\documentclass{appolb}
\usepackage[utf8]{inputenc}
\usepackage{url}
\usepackage{amsmath}
\usepackage{color}
\usepackage{graphicx}
\usepackage{multicol}
\usepackage{enumerate}
\usepackage{verbatim}
\usepackage{setspace}
\usepackage{cite}

\usepackage{xspace}
\usepackage{amsmath}

\newcommand{\NASixtyOne}{NA61\slash SHINE\xspace}
\newcommand{\eV}{\ensuremath{\mbox{e\kern-0.1em V}}\xspace}
\newcommand{\GeV}{\ensuremath{\mbox{Ge\kern-0.1em V}}\xspace}
\newcommand{\MeV}{\ensuremath{\mbox{Me\kern-0.1em V}}\xspace}
\newcommand{\GeVc}{\ensuremath{\mbox{Ge\kern-0.1em V}\!/\kern-0.07em c}\xspace}
\newcommand{\GeVcc}{\ensuremath{\mbox{Ge\kern-0.1em V}\!/\kern-0.07em c^2}\xspace}
\newcommand{\AGeV}{\ensuremath{A\,\mbox{Ge\kern-0.1em V}}\xspace}
\newcommand{\AGeVc}{\ensuremath{A\,\mbox{Ge\kern-0.1em V}\!/\kern-0.07em c}\xspace}
\newcommand{\MeVc}{\ensuremath{\mbox{Me\kern-0.1em V}\!/\kern-0.07em c}\xspace}

\newcommand{\pT}{\ensuremath{p_\text{T}}\xspace}

\newcommand{\km}{\ensuremath{\textup{K}^-}\xspace}
\newcommand{\kp}{\ensuremath{\textup{K}^+}\xspace}

\newcommand{\CernVM}{\textsc{Cern\-\kern-0.05emVM}\xspace}

\begin{document}

\title{Recent results and future of the \NASixtyOne strong interactions program%
\thanks{Presented at the XXXV Mazurian Lakes Conference on Physics, Piaski, Poland, September 3-9, 2017}
}

\author{Bartosz Łysakowski for the \NASixtyOne Collaboration
\address{ Institute of Physics, University of Silesia, Poland}
}
\maketitle
  \begin{abstract}
\NASixtyOne is a fixed target experiment at the CERN Super-Proton-Synchrotron. The main
goals of the experiment are to discover the critical point of strongly interacting matter and
to study
the properties of the onset of deconfinement. In order to reach these goals the collaboration studies hadron
production properties in nucleus-nucleus, proton-proton and proton-nucleus interactions.\\
 \indent In this talk, recent results on particle production in p+p interactions, as well as Be+Be and Ar+Sc collisions in the SPS energy
range, are reviewed. The results are compared with available world data. The future of the \NASixtyOne scientific program is also presented. 
  \end{abstract}
  
\section{Introduction}
The \NASixtyOne experiment performs a unique two-dimensional scan of the
phase diagram of strongly interacting matter.  The main goals are the study
of the properties of the onset of deconfinement by measurements of hadron
production and the search for the critical point of strongly interacting
matter by measuring event-by-event fluctuations.  Measurements are performed
in a wide beam momentum range (from 13 up to 150/158\AGeVc) and for various
systems (p+p, p+Pb, Be+Be, Ar+Sc, Xe+La and Pb+Pb).  The program is
motivated by the discovery of the onset of deconfinement in Pb+Pb collisions
at 30\AGeVc by the NA49 experiment\cite{calt,afan}.

\NASixtyOne is a fixed target experiment at the CERN SPS \cite{abgra}. The detection system is based
on eight Time Projection Chambers (TPC) providing acceptance in the full forward hemisphere, down
to \pT = 0. The TPCs allow for tracking, momentum and charge reconstruction as well as measurement of mean energy
loss per unit path length. Time of Flight (ToF) walls provide additional particle identification
by measuring particles mass. 
The Projectile Spectator Detector (PSD), a zero-degree calorimeter, allows
selecting collision centrality based on the measurement of forward energy. 
The full detector system of the \NASixtyOne experiment is presented in
Fig.~1.

\begin{figure}[htb]
\centerline{
\includegraphics[scale=0.37]{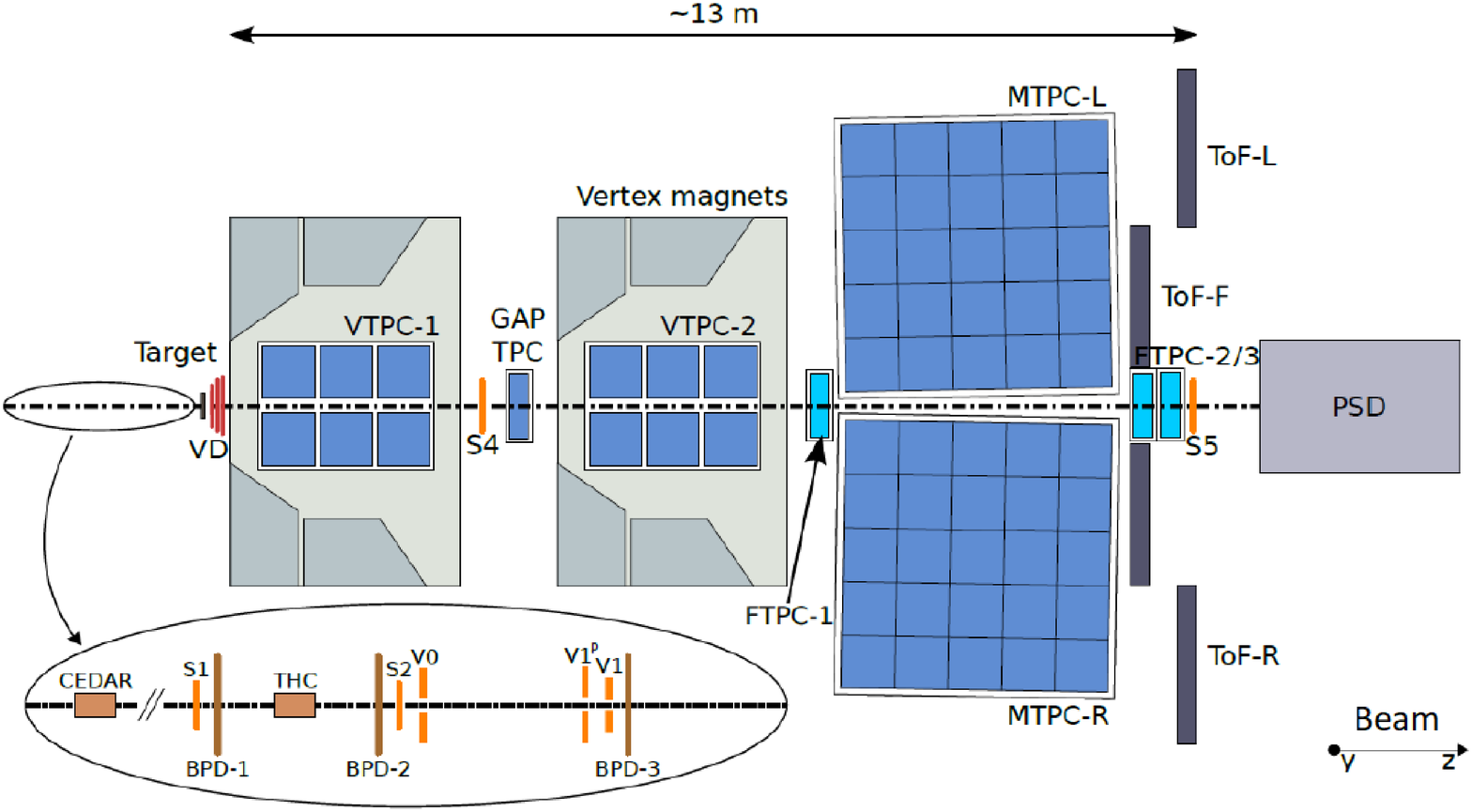}}
\caption{\NASixtyOne detector system}
\label{Fig:kink}
\end{figure}

\section{Onset of deconfinement}
One of the main purposes of NA61/SHINE is to study the phase transition
between hadron gas (HG) and quark-gluon plasma (QGP).  The measurements
concentrate on observables for which the Statistical Model of the Early
Stage (SMES) predicts characteristic signals \cite{SMES}.

Mean multiplicities of all pions, $\langle\pi\rangle$, normalized to the average number
of wounded nucleons, $\langle$W$\rangle$, are shown in Fig.~2.  The results were compared
with world data from other experiments \cite{ahle, blobel, golo}.  At higher
SPS energies the slope of the energy dependence is larger for the heavy
systems
(Pb+Pb, Ar+Sc) than for the light ones (p+p, Be+Be).  SMES predicts an
increase of the slope in the quark-gluon plasma due to a larger number of
degrees of freedom.  All A+A data come from central collisions.\\
\begin{figure}[!htb]
\centerline{
\includegraphics[width=6cm]{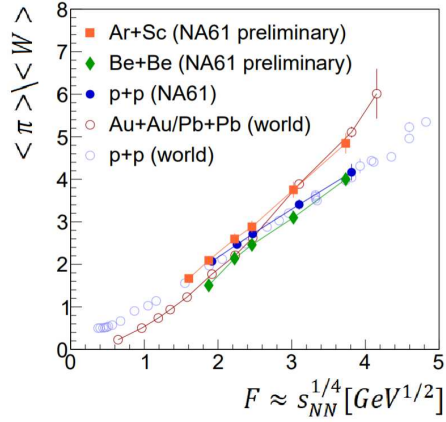}}
\caption{Energy dependence of the ratio $\langle\pi\rangle$/$\langle$W$\rangle$ of pion multiplicity to the number of wounded nucleons.}
\label{Fig:kink}
\end{figure}

Fig.~3 presents the multiplicity ratio of charged kaons to pions at
mid-rapidity.  Fig.~4 shows the energy dependence of the inverse slope
parameter of transverse mass spectra of charged kaons.  \NASixtyOne results
on p+p interactions \cite{szymon} and Be+Be collisions were compared with
results from central Pb+Pb collisions from NA49 \cite{calt,afan} and other
experiments \cite{15, 16, 17, 18, 19, 21}.\\

\begin{figure}[!htb]
\centerline{
\includegraphics[width=1.\textwidth]{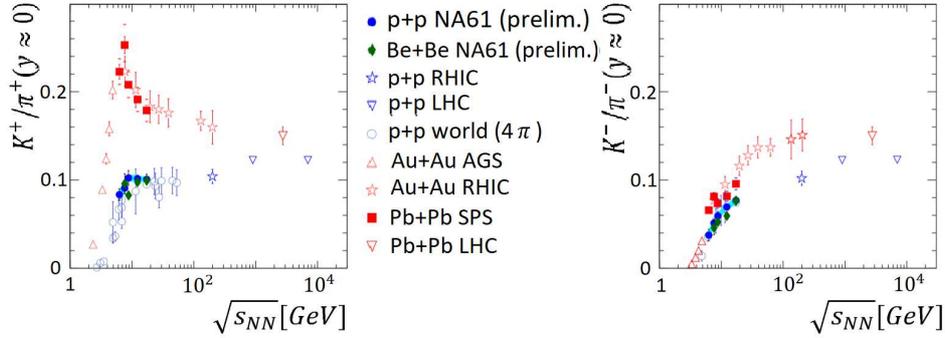}}
\caption{Energy dependence of the \kp (left) and \km (right) multiplicity divided by the corresponding charged $\pi$ multiplicity at
mid-rapidity.}
\label{Fig:kink}
\end{figure}

A plateau visible in the energy dependence of the inverse slope parameter in
Fig.~4,  and peaks seen in Fig.~3 (left panel) for Pb+Pb and Au+Au
collisions in the SPS energy range, were predicted by the SMES model as
signatures of the onset of deconfinement.  In the SPS energy range the
NA61/SHINE results on p+p interactions exhibit a qualitatively similar
energy dependence (step) for the inverse slope parameter, and a step instead
of a peak in the kaon-to-pion ratio.  Thus some properties of hadron
production previously attributed to the onset of deconfinement in heavy-ion
collisions are present also in p+p interactions.  Surprisingly, while the
inverse slope parameter in Be+Be collisions lies slightly above that  in
p+p interactions, the values of the charged-kaon-to-pion ratio are very
close in Be+Be and p+p reactions.\\

\begin{figure}[!htb]
\centerline{
\includegraphics[width=1.\textwidth]{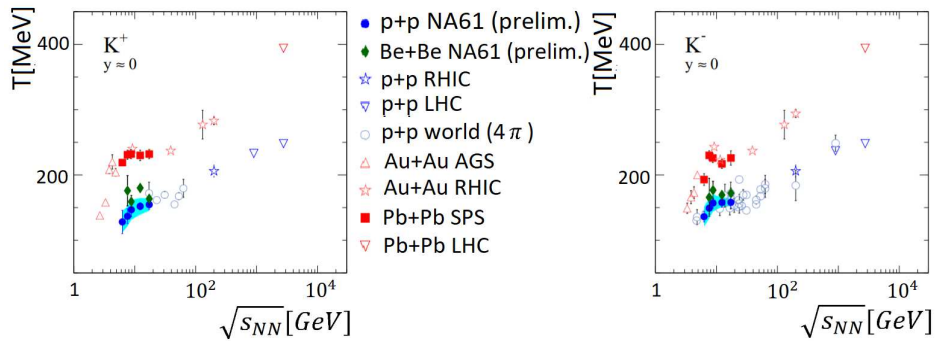}}
\caption{Energy dependence of the inverse slope parameter of the transverse mass spectra of \kp (left) and \km (right).}
\label{Fig:kink}
\end{figure}
\section{Search for the critical point}
\NASixtyOne uses the strongly intensive measures $\Sigma[P_T, N]$ and
$\Delta[P_T, N]$ to study transverse momentum and multiplicity fluctuations
\cite{22}.  Within the Wounded Nucleon Model, they depend neither on the
number of wounded nucleons (W) nor on fluctuations of W.  Moreover, in the
Grand Canonical Ensemble, they do not depend on volume and volume
fluctuations.  $\Sigma[P_T, N]$ and $\Delta[P_T, N]$ have two reference
values, namely, they are equal to zero in case of no fluctuations and one in
case of independent particle production.  Fig.~5 shows 
deviations of $\Sigma[P_T, N]$  and $\Delta[P_T, N]$ from unity
(independent particle model) that are smoothly growing  with energy, which may be due to the increasing
azimuthal acceptance.  So far, there are no prominent structures in the
NA61/SHINE data which could be related to the critical point (see the
expected signal from the critical point in Fig.  5, upper panel).

\begin{figure}[htb]
\centering
\includegraphics[width=0.5\textwidth]{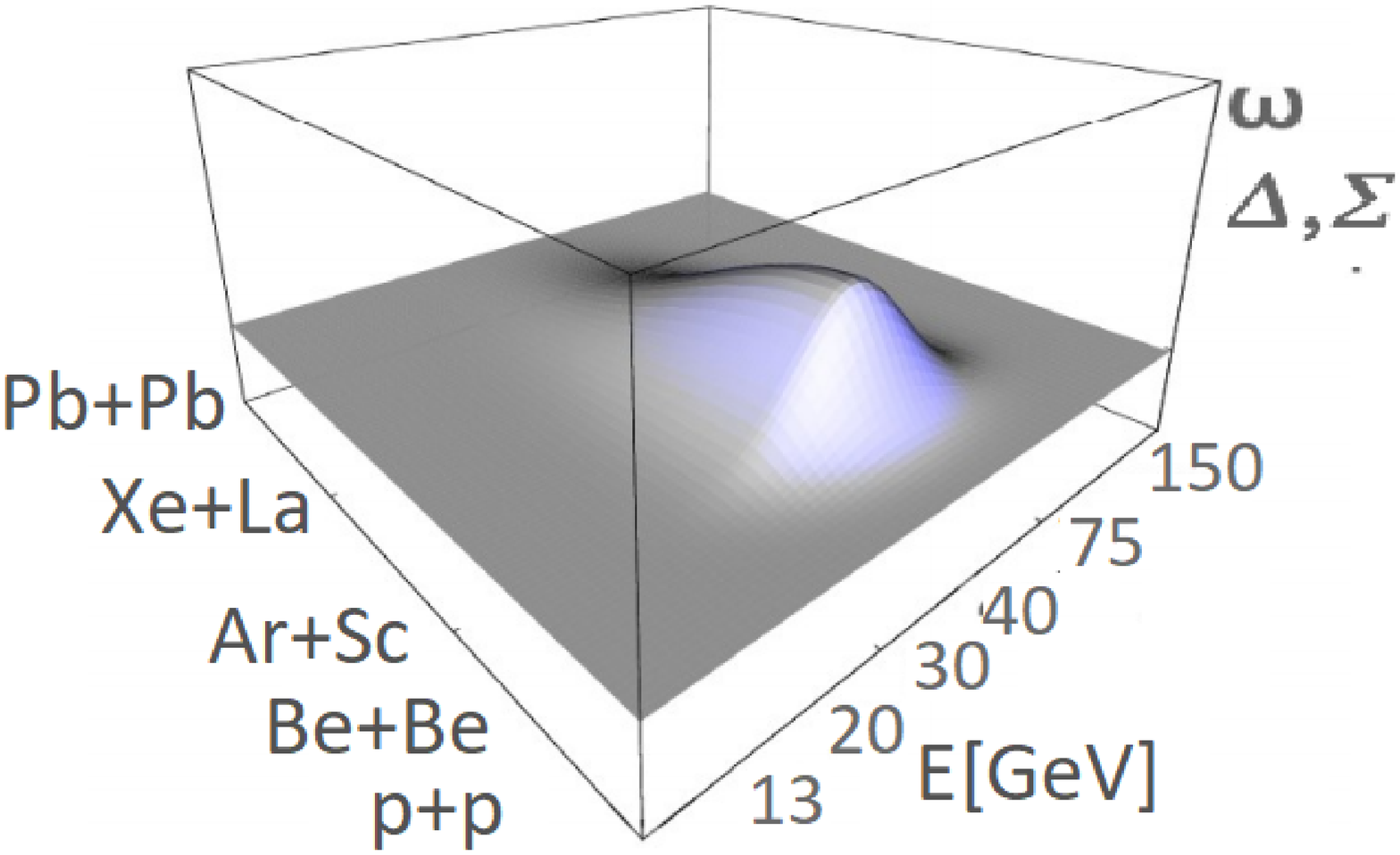}\\
\includegraphics[width=1.\textwidth]{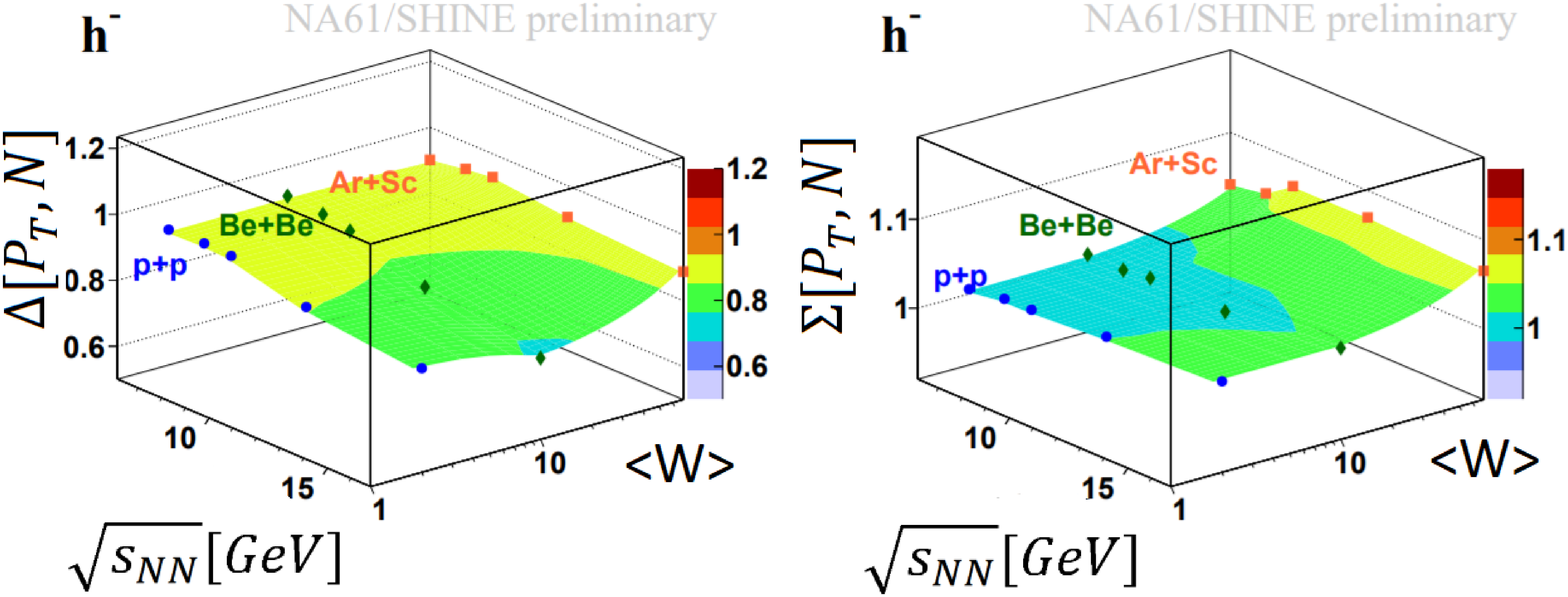}
\caption{{\em Upper panel}: a hill expected in an energy and system size scan for a fluctuation observable.
{\it Lower left
panel}: $\Delta[P_T, N]$  {\it Lower right
panel}: $\Sigma[P_T, N]$.Both lower panels are for negatively charged hadrons for 5\% of the most central collisions.}
\label{Fig:kink}
\end{figure}
%\section{$\Phi, \Lambda$ production}
\newpage
\section{NA61/SHINE detector upgrade}
This year \NASixtyOne started a pilot program of open charm measurements with the new Small Acceptance Vertex Detector (SAVD).\\

The construction of a high-resolution vertex detector was mostly motivated by the importance and the possibility
of the first direct measurements of open charm meson production in heavy-ion collisions at SPS
energies. In the first step a limited-acceptance version of the detector, named SAVD, was built. This device covers 35\% of the acceptance of the final version, and is foreseen to operate beyond 2020. The constructed SAVD is based on MIMOSA-26AHR sensors developed in IPHC Strasbourg.\\

A detector upgrade is planned during the long shutdown LS2 at CERN during the years 2019–2020: the readout speed will be increased to 1 kHz and
the Large Acceptance Vertex Detector will be constructed.\\

The upgraded detector will allow a high-statistics beam momentum scan with Pb+Pb
collisions for precise measurements of D-meson production and multi-strange hyperon production in 2021–2024.
%\begin{figure}[ht]
%\centerline{
%\includegraphics[width=5cm]{vertex.eps}}
%\caption{Visualization of SAVD}
%\label{Fig:kink}
%\end{figure}

\section{Summary}
This contribution briefly discusses recent results from the NA61/SHINE energy and system size scan performed to study the onset of deconfinement and search for the critical point of strongly interacting matter. Results on produced particle multiplicity and fluctuations were presented.

The measured fluctuations show no indication of a critical point. Still, such features may be revealed in future results on Xe+La and Pb+Pb collisions.

In addition, the extension of the NA61/SHINE program, such as the detector
system upgrade and plans to measure precisely open charm in 2021–2024, were
discussed in this contribution.

\section*{Acknowledgments}
This work was partially supported by the National Science Centre, Poland (grant 2015/18/M/ST2/00125).

\end{document}